# Connection Probability for Random Graphs with Given Degree Sequence*

Xu XinPing+

*Institute of Particle Physics, HuaZhong Normal University, Wuhan 430079, China*

*Abstract: Recently, the classical configuration model for random graphs with given degree distribution has been extensively used as a null model in contraposition to real networks with the same degree distribution. In this paper, we briefly review the applications of this model and derive analytical expression for connection probability by the expanding coefficient method. We also use our expanding coefficient method to obtain the connection probability for the directed configuration model.*

## 1. Introduction

Many natural and man-made complex systems can be fruitfully represented and studied in items of networks or graphs [1]. A random graph is a collection of points, or vertices, with lines, or edges, connecting pairs of them at random. The study of random graphs has a long history. Starting with the influential work of Erdo̎s and Ré nyi in the 1950s and 1960s [2－4], random graph theory has developed into one of the mainstays of modern discrete mathematics. However, the classical random graphs (or Poisson random graphs) have severe shortcomings as models of the real-world phenomena since it is a static model yielding small-world networks with a Poisson degree distribution. In fact, despite the network diversity, most of real networks have a scale-free topological structure in which the degree follows a power law distribution [5].

Given the substantial differences in the structure of degree distribution between the classical random graphs and real networks, it would clearly be useful if we could generalize the classical random graphs to incorporate arbitrary non-Poisson degree distribution. The method for doing this is to use the configuration model with given degree sequence, which has been put forward first by Bender and Canfield [6]. Since the 1970s the configuration model has been studied by a number of authors [7-9]. A variety of properties including the position of the phase transition at which a giant component first forms, the mean component size, the size of the giant component if there is one, and the average vertex-vertex distance within a graph etc are studied by Molloy and Reed [8, 10]. The configuration model has been one of the most successful algorithms proposed for network formation [8, 10 ], the relevance of the algorithm relies on its ability to generate random networks with a preassigned degree sequence—taken from a given degree distribution—at the user's discretion while maximizing the network's randomness at all other respects.

As a matter of fact, the configuration model has been extensively used as a null model in contraposition to real networks with the same degree distribution. For instance, to find motifs in biological networks, as well as networks from engineering, ecology, and other fields, we compare small subgraphs that occur in the real network to an ensemble of randomized networks with the same degree sequence [11,12]. The configuration model with given degree sequence is also applied to the problems of network robustness and of epidemics spreading on contact networks as a standard model for the dynamical process taking place on complex networks [13]. The configuration model for random graphs is also an ensemble model, which has been studied from the principles of equilibrium statistical mechanics [14, 15].

In this paper, we will study the traditional configuration model from the combinatorics. We will derive the two-node connection probability of this model by the expanding coefficient method. The paper is organized as follows. Section 2 gives a general description for the configuration model with given degree sequence. In section 3, we find analytical expression for the connection probability by using the expanding coefficient method. In Section 4, we apply our method to the directed case. Conclusions are given in the last part, Sec. 5.

* Supported by NSFC under projects 10375025, 10275027 and by the MOE under project CFKSTIP-704035.
+Email: xuxp@ihep.ac.cn

## 2. The configuration model

The configuration model is defined in the following way [16]. We specify a degree distribution $p(k)$, such that $p(k)$ is the fraction of vertices in the network having degree $k$. We choose a degree sequence, which is a set of $N$ values of the degrees $k_i$ of vertices $i = 1 \ldots N$, from this distribution. We can think of this as giving each vertex $i$ in our graph $k_i$ stubs sticking out of it, which are the ends of edges-to-be. Then we choose pairs of stubs at random from the network and connect them together to make complete edges. It is straightforward that this process generates every possible topology of a graph with the specified degree sequence.

In configuration model, the total number of nodes is fixed to $N$ and degrees of all nodes $i=1, 2\ldots, N$ create a specific degree sequence $\{k_i\}$. Until now, nothing has been said about the connections between nodes. In fact, this process produces a microcanonical ensemble of graphs with the desired degree sequence, which seems to be more familiar to physicists. Despite the broad application of the configuration model in the literature, an exact calculation for the two-node connection probability is still absent. In the next section, we will give answer to this question by the expanding coefficient method.

## 3. The connection probability

To calculate the connection probability of the two arbitrary nodes $m$ and $n$, we need to calculate the size of graph ensemble (or the number of single graphs) produced by the configuration model. To address this question, we construct a polynomial $f_L(x_1, x_2, \ldots x_N)$ to embody the graph ensembles in which only the total number of edges L is fixed. $f_L(x_1, x_2, \ldots x_N)$ is defined as follows

$$f_L(x_1, x_2, \ldots x_N) = (\sum_i^N x_i)^{2L} = \sum_{k_1, k_2, \ldots k_N} \frac{2L!}{\prod_i k_i!} \prod_i x_i^{k_i} \tag{1}$$

Where $x_i$ stands for one stub sticking out of node $i$ and $x_i^{k_i}$ stands for $k_i$ stubs sticking out of node $i$. Therefore, $f_L(x_1, x_2, \ldots x_N)$ presents the all the graph ensembles which are composed of 2L stubs (or L edges). The number of single graphs with the specified degree sequence $\{k_i\}$ is given by the expanding coefficients of the item $\prod_i x_i^{k_i}$ in $f_L(x_1, x_2, \ldots x_N)$. Thus the size of the graph ensemble of the configuration model $m$ is

$$m = \frac{2L!}{\prod_i k_i!} \tag{2}$$

Now we consider the ensemble in which node $m$ and $n$ are not connected. Such an ensemble can be represented by another polynomial

$$f_L^{\overline{mn}}(x_1, x_2, \ldots x_N) = [(\sum_i^N x_i)^2 - 2x_m x_n]^L = \sum_{i=0}^{L} \frac{L!}{(L-i)!i!}(\sum_j^N x_j)^{2(L-i)}(-2x_m x_n)^i \tag{3}$$

The contribution from connection between $m$ and $n$ has been subtracted by the item $2 x_m x_n$. Similarly, the number of single graphs with the same desired degree sequence $\{k_i\}$ is given by the expanding coefficients of the item $\prod_i x_i^{k_i}$ in $f_L^{\overline{mn}}(x_1, x_2, \ldots x_N)$. Consequently, the size of the graph ensemble $m^{\overline{mn}}$ with the same degree sequence in which $m$ and $n$ not connected is written as

$$m\overline{^{mn}} = \frac{2L!}{\prod_i k_i!}(1 - \frac{k_m k_n}{2L-1} + \frac{k_m k_n (k_m-1)(k_n-1)}{(2L-1)(2L-3)} - \ldots) \qquad (4)$$

Hence, the ensemble size (number of single graphs) in which *m* and *n* are connected yields

$$m^{mn} = m - m\overline{^{mn}} \qquad (5)$$

The connection probability $p_{mn}$ is given as the ratio between $m^{mn}$ and $m$

$$p_{mn} = \frac{m^{mn}}{m} = \frac{k_m k_n}{2L-1} - \frac{k_m k_n (k_m-1)(k_n-1)}{(2L-1)(2L-3)} + \ldots \qquad (6)$$

Thus we derived the connection probability for random graphs with given degree sequence. In particular, if the first item satisfies $k_m k_n /(2L-1) << 1$, the contribution from higher order items of Eq.(6) is neglectable. Hence, we have

$$p_{mn} = \frac{k_m k_n}{2L-1} \approx \frac{k_m k_n}{2L} \qquad (7)$$

The connection probability (7) is proportional to the product of the degrees of its endpoints, which is precisely the model of Chung and Lu (CL model) [9], but is obtained under the approximation "sparse limit" or the "classical limit" $k_m k_n /(2L-1) << 1$. Eq.(7) has significant physical meaning, e.g., it can be interpreted as justifying the conventional rule of linear preferential attachment in growing networks [17].

We would like to point out that, the number of single graphs of the ensemble with given degree sequence calculated in this paper, includes the graphs with self connections or multiple connections. In order to present this issue, we consider the self connections of a certain node *s* in the ensemble. The number of single graphs of the ensemble with given degree sequence without self connections of node *s* is given by the expanding coefficients of the item $\prod_i x_i^{k_i}$ in $[(\sum_i x_i)^2 - x_s^2]^L$, thus the probability of node *s* with self connections $p(s \leftrightarrow s)$ reads

$$p(s \leftrightarrow s) = \frac{k_s(k_s-1)}{2(2L-1)} - \frac{L(L-1)}{2!} \cdot \frac{k_s(k_s-1)(k_s-2)(k_s-3)}{2L(2L-1)(2L-2)(2L-3)} + \ldots \qquad (8)$$

Which turns to $p(s \leftrightarrow s) = \frac{k_s(k_s-1)}{2(2L-1)}$ under the first order approximation. This formula indicates that the occurrence of self connections in a particular node is nontrivial.

**4. The directed case.**

Directed networks are more complex than their undirected counterparts. In this section, we extend the above methodology to the directed case. For a start, each vertex in a directed network has two degrees, an in-degree, which is the number of edges that point into the vertex, and an out-degree, which is the number pointing out. Thus, there are two degree distributions, in-degree sequence $\{k_i^{in}\}$ and out-degree sequence $\{k_i^{ou}\}$ to describe the directed configuration model. We note that $\sum_i k_i^{in} = \sum_i k_i^{ou}$, as must be the case for all directed graphs, since every edge on such a graph must both start and end at exactly one vertex.

Consider the graph ensemble in which the total number of directed edges L is fixed, we can construct a polynomial $f_L(\vec{x}_1, \vec{x}_2, \ldots \vec{x}_N; \bar{x}_1, \bar{x}_2, \ldots \bar{x}_N)$ as follows

$$f_L(\vec{x}_1, \vec{x}_2, \ldots \vec{x}_N; \bar{x}_1, \bar{x}_2, \ldots \bar{x}_N) = (\sum_i^N \vec{x}_i)^L (\sum_i^N \bar{x}_i)^L \tag{9}$$

Where $\vec{x}_i$ and $\bar{x}_i$ stands for the stubs sticking in and out of vertex *i*. The number of single graphs which have the given degree sequence $\{k_i^{in}\}$ and $\{k_i^{ou}\}$ is

$$w = \frac{L!}{\prod_i k_i^{in}!} \bullet \frac{L!}{\prod_i k_i^{ou}!} \tag{10}$$

Similarly, the ensemble in which there is no connection $m \to n$ can be represented by the following polynomial

$$f_L^{\overline{m \to n}}(\vec{x}_1, \vec{x}_2, \ldots \vec{x}_N; \bar{x}_1, \bar{x}_2, \ldots \bar{x}_N) = [(\sum_i^N \vec{x}_i)(\sum_i^N \bar{x}_i) - \bar{x}_m \vec{x}_n]^L \tag{11}$$

The contribution from connection $m \to n$ has been removed by the item $\bar{x}_m \vec{x}_n$. The number of single graphs with the same desired degree sequence $\{k_i^{in}\}$ and $\{k_i^{ou}\}$ is given by the expanding coefficients of the item $\prod_i \vec{x}_i^{k_i^{in}} \prod_i \bar{x}_i^{k_i^{ou}}$ in $f_L^{\overline{m \to n}}(\vec{x}_1, \vec{x}_2, \ldots \vec{x}_N; \bar{x}_1, \bar{x}_2, \ldots \bar{x}_N)$, which yields

$$w^{\overline{m \to n}} = \frac{L!}{\prod_i k_i^{in}!} \bullet \frac{L!}{\prod_i k_i^{ou}!} (1 - \frac{k_n^{in} k_m^{ou}}{L} + \frac{1}{2!} \bullet \frac{k_m^{ou} k_n^{in}}{L} \bullet \frac{(k_m^{ou}-1)(k_n^{in}-1)}{L-1} - \ldots) \tag{12}$$

We can also obtain the connection probability $p(m \to n)$ in a natural fashion analogous to the undirected case. In the sparse limit $k_n^{in} k_m^{ou} / L \ll 1$, the probability of *m* connecting *n* is

$$p(m \to n) \approx \frac{k_n^{in} k_m^{ou}}{L} \tag{13}$$

This gives similar results to Eq.(7), in which the connection probability is proportional to the product of the degrees of its endpoints. After a simple algebra, one can confirm that the expected out-degree $<k_m^{ou}>$ (or in-degree) is equal to its given out-degree $k_m^{ou}$ (or in-degree) of vertex *m*,

$$<k_m^{ou}> = \sum_n p(m \to n) = k_m^{ou} \frac{\sum_n k_n^{in}}{L} = k_m^{ou} \tag{14}$$

This is consistent with our intuition since each single graph has the same degree sequence and the ensemble mean of the degrees is just the given degrees.

**5. Conclusions.**

In conclusion, we have briefly reviewed the applications of random graphs with given degree sequence and find analytical expression for connection probability by the expanding coefficient method. We also use our expanding coefficient method to obtain the connection probability for the directed configuration model. Particularly, the expanding coefficient method introduced by us provides us a powerful tool in our calculation. We hope the expanding coefficient approach adopted in this paper can shed much light on the understanding of graph ensemble and will be useful in the future research.